\documentclass[12pt,a4paper]{article}
\usepackage{dcolumn}
\usepackage{bm}
\usepackage{amssymb}
\usepackage{tabularx}
\usepackage{booktabs}
\usepackage{amsmath}
\usepackage{subcaption}
\usepackage{diagbox} 
\usepackage{hyperref}
\usepackage{graphicx}
\usepackage{lmodern}
\usepackage[left=2cm,right=2cm,top=2cm,bottom=2cm]{geometry}
\usepackage{authblk}
\title{Physics-Guided Neural Networks for Constructing Nucleon-Nucleon Inverse Potentials}
\author[1]{Ayushi Awasthi}
\author[2]{Anil Khachi}
\author[3]{M.R. Ganesh}
\author[1]{O.S.K.S. Sastri}
\affil[1]{Department of Physics and Astronomical Sciences, Central University of Himachal Pradesh, Dharamsala, Bharat (India)}
\affil[2]{ Department of Applied Sciences,Chandigarh Group of Colleges, Jhanjeri, Mohali, Punjab,140307, India (Bharat) }
\affil[3]{ Applied Materials India Private Limited, Bengaluru, Karnataka,560066, India (Bharat)}
\begin{document}
\maketitle
\begin{abstract}
We propose a physics-guided neural network (PGNN) framework for constructing nucleon-nucleon inverse potentials based on inverse scattering theory. The framework integrates the Phase Function Method (PFM) with a two-stage supervised multi-layer perceptron (MLP) model to extract the optimal parameters of the Malfliet-Tjon (MT) potential from sparse phase-shift data. A synthetic dataset of phase shifts is generated by solving the phase equation for angular momentum $\ell$ = 0, using the fifth-order Runge-Kutta method, ensuring physically consistent training data. The first neural network predicts the attractive potential strength, $\tilde{V}_A$, while the second estimates the repulsive strength, $\tilde{V}_R$. The optimal range parameter, $\mu$, is obtained through error minimization between predicted and expected phase shifts, thereby enhancing both stability and accuracy compared to conventional inversion techniques. The PGNN framework is validated for $^1S_0$ state of neutron-proton (n-p), proton-proton (p-p), and neutron-neutron (n-n) scattering at low energies. The constructed inverse potentials accurately reproduce phase shifts reported in the literature and exhibit the expected features of nucleon-nucleon interactions, including a short-range repulsive core and an intermediate-range attractive well, with the n-p system showing the deepest potential minimum due to stronger binding. These results demonstrate that the proposed PGNN framework provides an efficient and accurate approach for constructing nuclear potentials, effectively bridging machine learning techniques with quantum scattering theory.
 
\vspace{0.5cm}
\textbf{Keywords:}
Neural Networks, Inverse scattering, Phase Function Method, physics-guided learning,Multi-layer perceptron, nucleon-nucleon interaction.
\end{abstract} 

\section{Introduction}
Understanding the nucleon-nucleon (NN) interaction remains one of the central challenges in nuclear physics, as it forms the foundation for describing nuclear structure~\cite{PhysRevC.108.044617}, reaction dynamics~\cite{PhysRevC.111.054621}, and astrophysical processes such as nucleosynthesis and neutron star properties~\cite{Lu2021SRC}. Accurate knowledge of the NN potential is essential for reliable predictions of binding energies, scattering observables, and the equation of state of nuclear matter \cite{PhysRevC.104.054001}. Over the past several decades, significant efforts have been devoted to constructing high-precision NN interaction models, leading to a wide variety of theoretical approaches \cite{Naghdi2014}.  
Traditional phenomenological potentials, such as Argonne $V_{18}$ ~\cite{PhysRevC.51.38}, CD-Bonn~\cite{PhysRevC.63.024001}, and Nijmegen~\cite{PhysRevC.49.2950}, have achieved remarkable success in reproducing nucleon-nucleon scattering phase shifts and deuteron properties with high accuracy. These models are typically based on fitting a large number of free parameters (usually around 40–50) to experimental data, effectively encoding the physics within empirical forms of the potential. However, such approaches have several important limitations. First, they lack predictive power when extrapolated to energy or density regimes where experimental data are scarce or absent. Second, their dependence on extensive parameter fitting often leads to ambiguities in interpreting the underlying physics. Finally, because these models are computationally demanding, their direct application to large-scale nuclear many-body calculations is challenging. \\
In recent years, chiral effective field theory (\(\chi\)EFT) has emerged as a systematic and theoretically consistent framework for constructing NN interactions grounded in the symmetries of quantum chromodynamics (QCD) \cite{MACHLEIDT20111}. By expanding the nuclear interaction in powers of momentum over a breakdown scale, \(\chi\)EFT provides a hierarchy of contributions and a natural way to estimate theoretical uncertainties. While \(\chi\)EFT-based potentials have significantly improved the description of low-energy nuclear phenomena, they also introduce new challenges. In particular, predictions depend on the choice of resolution scale (cutoff) and the regularization scheme, leading to non-negligible model dependencies and uncertainties in many-body nuclear calculations \cite{Furnstahl_2015}. Moreover, despite the systematic nature of \(\chi\)EFT, achieving a fully accurate description of NN observables across different scales remains an open problem. \\

The rapid development of machine learning (ML) has recently opened promising avenues for addressing some of these limitations. ML methods are capable of learning complex mappings between observables and model parameters, even in the presence of incomplete or noisy data \cite{JOHNSTON2022100231}. Notably, Wen \textit{et al.}~\cite{PhysRevLett.133.252501} demonstrated that generative modeling techniques can be successfully applied to the NN interaction problem. By training on existing \(\chi\)EFT-derived potentials at different orders and resolution scales, their generative model was able to construct continuous families of NN interactions and produce high-quality scattering phase shifts across a broad parameter space. This work highlights the potential of combining physics-based models with data-driven techniques to systematically propagate theoretical uncertainties and accelerate nuclear many-body calculations. \\
Parallel to these developments, inverse scattering theory provides an alternative approach for constructing NN interaction potentials directly from experimental data \cite{Ismailov2025}. Among the various inversion techniques, the Phase Function Method (PFM) \cite{Calogero1967,Babikov1967} has emerged as a widely used and effective approach. By reformulating the second-order Schrödinger equation into a first-order Riccati-type equation, PFM allows efficient computation of phase shifts and potentials without requiring explicit solutions of the full wave function \cite{Khachi2023}. Previously, we developed a framework combining the Phase Function Method with a reference potential approach ~\cite{Awasthi_2024}. In this work, three Morse functions were smoothly joined to represent the short-range repulsive core, intermediate-range attractive well, and long-range asymptotic behavior of the potential. The model parameters were optimized using a Genetic Algorithm (GA) \cite{Katoch2021}, providing a robust global search mechanism to minimize deviations between computed and expected phase shifts. For charged-particle systems, the third Morse component was modified to incorporate Coulomb effects. This methodology was successfully applied to several few-body and light-nucleus scattering systems, including $\alpha$-$^3\mathrm{He}$, $\alpha$-$^3\mathrm{H}$ \cite{Kant_2025}, neutron-deuteron, proton-deuteron \cite{AWASTHI2025109800}, $\alpha$-deuteron \cite{sharma2025_}, and $\alpha$-$^{12}\mathrm{C}$ scattering \cite{awasthi2025_GA}, demonstrating that carefully designed potentials combined with physics-based constraints and robust optimization can yield inverse potentials across a broad range of nuclear processes.However, the inverse scattering problem is inherently ill-posed: small uncertainties in the input phase shifts can lead to large variations in the constructed potentials. Furthermore, the limited availability of high-precision phase-shift data at low and intermediate energies poses significant challenges to constructing reliable inverse potentials \cite{Kress1989}.
These limitations have motivated the development of hybrid frameworks that integrate machine learning with physics-based inversion techniques \cite{Romualdi2021}. \\
One promising direction involves Physics-Guided Neural Networks (PGNNs), where neural network architectures are designed to incorporate physical constraints directly into the learning process \cite{RAISSI2019686}\cite{Zhao2024}. \\
Unlike purely data-driven models, PGNNs leverage governing equations, known symmetries, or synthetic datasets generated from well-established theoretical models, thus improving interpretability, enhancing predictive accuracy, and ensuring physically consistent results. Recent studies have shown the effectiveness of PGNNs in modeling complex quantum scattering processes \cite{Cuomo2022}, making them particularly suitable for constructing NN interaction potentials.
In this work, we present a novel PGNN-based framework for constructing nucleon-nucleon inverse potentials by combining PFM-based inverse scattering theory with modern machine learning techniques. Our approach uses the PFM to generate a large synthetic dataset of phase shifts by solving phase equation for $\ell$ = 0, using Runge-Kutta (5th) order method. These synthetic datasets provide the neural network with physically consistent training samples, ensuring that the predicted potentials inherently satisfy the underlying quantum scattering equations. We employ a two-stage supervised learning strategy using a multi-layer perceptron (MLP) architecture: the first network predicts the attractive strength parameter \(\tilde{V}_A\), while the second predicts the repulsive strength parameter \(\tilde{V}_R\). Additionally, the optimal range parameter \(\mu\) is determined by minimizing the discrepancy between the predicted and reference phase shifts, ensuring both accuracy and stability in potential construction. The proposed framework is designed to balance computational efficiency, physical consistency, and predictive flexibility. By embedding physical constraints directly into the machine learning architecture, our model provides reliable reconstructions of NN potentials even when experimental data are limited or noisy. Moreover, the integration of synthetic datasets generated from existing nuclear models ensures that the constructed potentials remain consistent with known quantum scattering properties while enabling systematic extrapolation to unexplored regimes. This work represents an important step toward unifying physics-driven inversion techniques with modern machine learning tools for high-precision modeling of nuclear forces.
\section{Methodology}\label{sec1}
In this section, we have discussed the physics-informed supervised learning \cite{Meng2025} framework for obtaining the scattering potential using the inverse scattering theory \cite{Balassa2022}. The work flow is composed of three main stages:
\begin{itemize}
    \item Physics guided synthetic data generation via Phase Function Method.
    \item Two stage neural-network training to map the model parameters with the input data.
    \item Inversion Process to find the optimal potential that produces the expected phase shift data.
\end{itemize}
This combination allows the model to learn from data that obey the underlying scattering equation thus improving interoperability and generalization. 
\subsection{Data Generation}
Expected phase shift data for nucleon–nucleon scattering \cite{PhysRevC.51.38} is typically very limited. Directly training a neural network on such limited datasets risks overfitting and poor extrapolation. To mitigate this, we generate a large synthetic dataset using the phase function method \cite{Calogero1967,Khachi2023}, ensuring that all training samples strictly satisfy the quantum mechanical scattering equation.
\subsubsection{Phase Function Method}
The phase function method (PFM) is an efficient approach to solve the inverse nuclear scattering problem by transforming the second-order Schrödinger equation into a first-order Riccati equation \cite{Babikov1967}. The phase function $\delta_\ell(k,r)$ represents the accumulated phase shift of the wave at radius $r$, accounting for the scattering effects of the potential $V(r)$. For elastic scattering, the phase equation is given by \cite{Khachi2023}:
\begin{equation}
\frac{d\delta_\ell(k,r)}{dr} = -\frac{V(r)}{k} \Big[\cos\delta_\ell\,\hat{j}_\ell(kr) - \sin\delta_\ell\,\hat{\eta}_\ell(kr)\Big]^2,
\end{equation}
where $k = \sqrt{\tfrac{2m E_{cm}}{\hbar^2}}$ is the wave number, V(r) is the interaction potential, and m is the reduced mass of the system. The center-of-mass energy $E_{cm}$ is related to the laboratory energy $E_{\ell ab}$ by
\begin{equation}
E_{cm} = \frac{m_T}{m_T + m_P} E_{\ell ab},
\end{equation}
where $m_T$ and $m_P$ are the masses of the target and projectile, respectively. The functions $\hat{j}_\ell(kr)$ and $\hat{\eta}_\ell(kr)$ are the Riccati-Bessel and Riccati-Neumann functions, while the Riccati-Hankel function of the first kind is defined as $\hat{h}_\ell(kr) = -\hat{\eta}_\ell(kr) + i\hat{j}_\ell(kr)$.  
For the $s$-wave case ($\ell=0$), $\hat{j}_0(kr)=\sin(kr)$ and $\hat{\eta}_0(kr)=-\cos(kr)$, reducing the equation to \cite{Calogero1967}:
\begin{equation}
\delta_0'(k,r) = -\frac{V(r)}{k} \sin^2\!\left[kr+\delta_0(k,r)\right].
\end{equation}
This non-linear equation is solved numerically using the Runge-Kutta 5th-order method with $\delta_\ell(0)=0$.
\subsubsection{Potential Model and Parameter Sampling}
In this work, the nucleon-nucleon ($NN$) interaction is modeled using the Malfliet-Tjon (MT) potential \cite{Malfliet1969,awasthi2025numerical}, which is widely employed to describe low-energy nuclear scattering. The potential is expressed as
\begin{equation}
    V(r) = \frac{V_R \, e^{-2\mu r} - V_A \, e^{-\mu r}}{r},
\end{equation}
where $V_R$ and $V_A$ denote the strengths of the repulsive and attractive components of the interaction, respectively, measured in $\mathrm{fm^{-2}}$, and $\mu$ represents the inverse range parameter in $\mathrm{fm^{-1}}$.  
\\
For the charged particle scattering (proton-proton ($p$--$p$) system), in addition to the nuclear interaction, the Coulomb repulsion must be taken into account. To incorporate this effect, we employ the Atomic Hulthén (AH) potential \cite{Awasthi2024}, defined as
\begin{equation}
    V_{\mathrm{AH}}(r) = V_0 \, \frac{e^{-r/a}}{1 - e^{-r/a}},
\end{equation}
where $V_0$ represents the potential strength and $a$ is the screening radius. These two parameters are related by the expression
\begin{equation}
    V_0 a = 2 K \eta,
\end{equation}
where $K$ is the relative momentum in the laboratory frame, and $\eta$ is the Sommerfeld parameter given by
\begin{equation}
    \eta = \frac{\alpha}{\hbar v},
\end{equation}
with $v$ being the relative velocity of the interacting particles at large separation and $\alpha = Z_1 Z_2 e^2$. Combining these relations yields
\begin{equation}
    V_0 a = \frac{Z_1 Z_2 e^2 m}{\hbar^2},
\end{equation}
where $m$ is the reduced mass of the two-body system.  
For the $p$-$p$ interaction, we have $Z_1 = Z_2 = 1$ and $m = m_p/2 = 469.136023\,\mathrm{MeV}/c^2$. Using $e^2 = 1.44\,\mathrm{MeV\,fm}$, we obtain
\begin{equation}
    V_0 a = 0.03472~\mathrm{fm^{-1}}.
\end{equation}
In the present study, we adopt a screening radius of $a = 5~\mathrm{fm}$, beyond which the Coulomb potential becomes negligibly small \cite{Awasthi2025}. This choice is particularly important in scattering calculations, as contributions from regions where the potential has effectively vanished do not significantly affect the interaction dynamics \cite{PhysRevC.109.064004}.
Since the total interaction depends on three parameters, we define these parameters within the ranges $V_R \in [-300, 300]~ fm^{-2}$, $V_A \in [-100, 100]~ fm^{-2}$, and $\mu \in [0.01, 4] fm^{-1}$. Using these ranges, we generated a total of $10,000$ unique potential configurations through uniform random sampling. For each sampled potential, the phase equation was solved using the fifth-order Runge-Kutta (RK-5) method for $^1S_0$ state of n-n, n-p and p-p at laboratory energies $E_{\mathrm{lab}} = [1,\, 5,\, 10,\, 25,\, 50,\, 100,\, 150,\, 200,\, 250,\, 300,\, 350]~\mathrm{MeV}$.
\\  In this manner, the data were generated by solving the quantum scattering equation, which were subsequently utilized for the training, testing, and validation processes \cite{PhysRevB.106.214307}.
\subsubsection{Data Normalization}
To ensure efficient model training and improve convergence, both the input features (phase shifts) and the target outputs were normalized to the range $[-1, 1]$ using the \texttt{MinMaxScaler} function from the \texttt{scikit-learn} library \cite{scikit-learn}. The normalization is performed according to the following formula:
\begin{equation}
    x_{\text{norm}} = 2 \cdot \frac{x - x_{\text{min}}}{x_{\text{max}} - x_{\text{min}}} - 1,
\end{equation}
where $x$ represents the original value, $x_{\text{min}}$ and $x_{\text{max}}$ denote the minimum and maximum values of the feature, respectively, and $x_{\text{norm}}$ is the normalized value within the range $[-1, 1]$.  Separate normalization scalers were applied for the input features and each set of output variables to preserve their individual data distributions. Normalization accelerates training convergence and helps prevent numerical instabilities in gradient-based optimization \cite{ioffe2015batch}.

\subsection{Neural Network Architecture}
\begin{figure}[h!]
    \centering
    \includegraphics[width=1.0\textwidth,height=0.7\textheight,keepaspectratio]{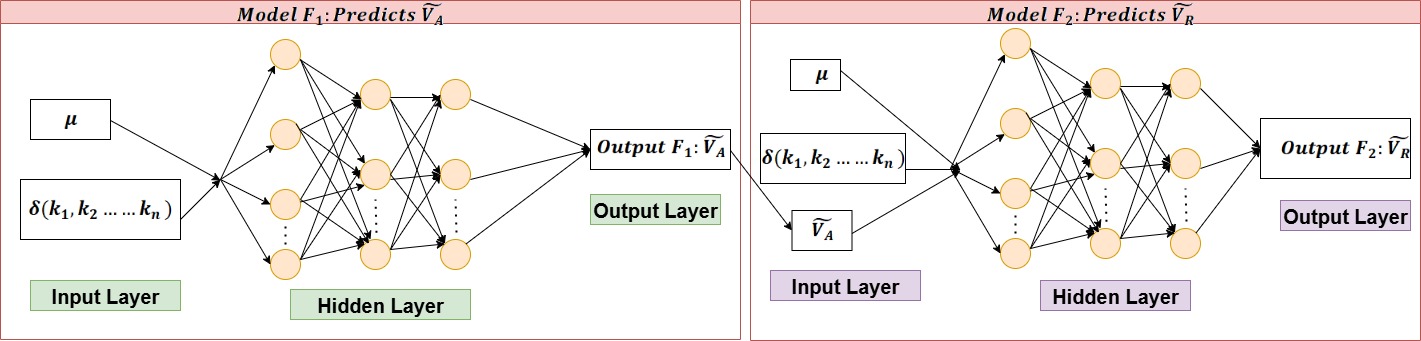} 
    \caption{Architecture of the proposed two-stage neural network framework. 
    Network $F_1$ predicts $\tilde{V}_A$ from the potential parameter $\mu$ and phase shifts $\delta(k_1), \delta(k_2), \dots, \delta(k_{11})$. 
    Network $F_2$ predicts $\tilde{V}_R$ using $\mu$, the estimated $\tilde{V}_A$, and the same set of phase shifts. 
    Both networks share an identical structure with three fully connected hidden layers containing $128$, $64$, and $64$ neurons, respectively, and use ReLU activation functions.}
    \label{Fig1_nn}
\end{figure}

To improve the stability and accuracy of the inversion process, the overall problem is reformulated as two sequential subproblems. Each subproblem is addressed independently using a dedicated neural network based on the Multilayer Perceptron (MLP) architecture \cite{article} as shown in Fig. \ref{Fig1_nn}. This decomposition helps to reduce the complexity of the mapping and allows each network to focus on learning a more specialized relationship between the inputs and outputs. In the first stage, represented by Network $F_1$, the objective is to estimate the attractive parameter $\tilde{V}_A$ using the parameter $\mu$ and the corresponding phase shifts $\delta(k_1), \delta(k_2), \dots, \delta(k_{11})$. Formally, this can be expressed as:
\[
\tilde{V}_A = F_1 \Big( \mu, \delta(k_1), \delta(k_2), \dots, \delta(k_{11}) \Big).
\]
This stage effectively captures the non-linear relationship between the input parameters and the attractive parameter $\tilde{V}_A$, which is subsequently used as an additional input for the second network.
In the second stage, represented by Network $F_2$, the goal is to predict the repulsive parameter $\tilde{V}_R$ using three sets of inputs: the parameter $\mu$, the previously estimated $\tilde{V}_A$ from $F_1$ model, and the same set of phase shifts $\delta(k_1), \delta(k_2), \dots, \delta(k_{11})$. The mapping can be formulated as:
\[
\tilde{V}_R = F_2 \Big( \mu, \tilde{V}_A, \delta(k_1), \delta(k_2), \dots, \delta(k_{11}) \Big).
\]
This sequential design ensures that the estimation of $\tilde{V}_R$ leverages both the raw input information and the intermediate representation learned in the first stage, improving prediction accuracy.
\subsubsection*{Network Architecture and Training Setup}
Both $F_1$ and $F_2$ share an identical architecture composed of three fully connected hidden layers with $128$, $64$, and $64$ neurons, respectively. The Rectified Linear Unit (ReLU) activation function is applied after each hidden layer to introduce non-linearity into the model, enabling it to learn complex mappings between the inputs and outputs \cite{Kulathunga2021}.
For optimization, the Adam optimizer \cite{kingma2015adam} is employed with hyperparameters $\beta_1 = 0.9$, $\beta_2 = 0.999$, and a learning rate of $\eta = 10^{-3}$. Adam was chosen due to its ability to adaptively adjust learning rates for each parameter, which facilitates faster convergence and better performance on regression tasks.
The Mean Squared Error (MSE) is used as the loss function \cite{Santosh2022}:
\[
\text{MSE} = \frac{1}{N} \sum_{i=1}^{N} \big( y_i - \hat{y}_i \big)^2,
\]
where $y_i$ represents the true value and $\hat{y}_i$ is the predicted value. MSE is well-suited for continuous-valued regression problems since it penalizes larger errors more strongly. To ensure proper model generalization, the dataset is divided into three subsets: $70\%$ for training, $20\%$ for validation, and $10\%$ for testing \cite{Shahrabadi2024}. The training set is used to optimize the model parameters, the validation set helps tune hyperparameters and avoid overfitting, and the testing set provides an unbiased evaluation of the model's performance.
The benefits of this two-stage approach can be summarized as follows:
\begin{enumerate}
\item \textbf{Reduced Learning Complexity:} By dividing the task into two smaller subproblems, each network focuses on learning a simpler and more specialized mapping.
\item \textbf{Improved Prediction Accuracy:} The intermediate parameter $\tilde{V}_A$ captures essential information, which enhances the accuracy of $\tilde{V}_R$ prediction.
\item \textbf{Enhanced Model Stability:} The staged structure reduces the risk of unstable convergence during training by simplifying the optimization process.
\item \textbf{Better Generalization:} By limiting the complexity of each subtask, the overall  model demonstrates improved generalization performance on unseen data.
\end{enumerate}
Thus, the proposed two-stage framework provides a more robust and reliable solution 
for the inversion problem, improving both the stability and accuracy of the final predictions.
\subsection{Inversion Procedure}
Once the neural networks $F_1$ and $F_2$ are trained, they are integrated into an inversion framework to estimate the optimal potential parameters for a given nucleon-nucleon (N-N) scattering system. 
The goal of the inversion is to determine the potential parameters $\mu$, $\tilde{V}_A$, and $\tilde{V}_R$ that best reproduce the expected phase shifts \cite{PhysRevC.51.38} by minimizing the error between simulated and expected data.
The inversion procedure consists of the following steps:
\begin{enumerate}
\item \textbf{Selection of Trial Parameter $\mu$:}  
A trial value of the potential parameter $\mu$ is selected within the physically allowed range $[0.01,\,4]$. The search is performed over this interval with a fine step size of $\Delta \mu = 0.01$ to ensure sufficient resolution.
\item \textbf{Prediction of $\tilde{V}_A$ Using $F_1$:} For each trial $\mu$, the trained neural network $F_1$ predicts the attractive component of the potential, $\tilde{V}_A$, using $\mu$ and the expected experimental phase shifts as inputs.
\item \textbf{Prediction of $\tilde{V}_R$ Using $F_2$:} The second network, $F_2$, predicts the repulsive component $\tilde{V}_R$ based on $\mu$, the estimated $\tilde{V}_A$, and the same expected phase shifts.
\item \textbf{Construction of the MT Potential:}  Using the current trial value of $\mu$ along with the predicted $\tilde{V}_A$ and $\tilde{V}_R$, the Malfliet--Tjon (MT) potential is constructed.
\item \textbf{Computation of Theoretical Phase Shifts:}  
The constructed MT potential is used to solve the relevant scattering equations, yielding the simulated phase shifts $\delta_{\text{sim}}(E_i)$ at the set of experimental energies $E_i$.
\item \textbf{Evaluation of the Mismatch via MSE:}  
To measure the difference between simulated and experimental phase shifts, the Mean Squared Error (MSE) is computed as:
\begin{equation}
\text{MSE} = \frac{1}{N} \sum_{i=1}^{N} \left[ \delta_{\text{sim}}(E_i) - \delta_{\text{exp}}(E_i) \right]^2,
\end{equation}
where $N$ is the total number of experimental energy points.
\item \textbf{Search for Optimal Parameters:}  
By sweeping over the entire range of $\mu$ with a fine step size $\Delta \mu = 0.01$, the MSE is evaluated for each trial value.  
The optimal $\mu^\ast$ is identified as the one that minimizes the MSE.
\item \textbf{Reporting the Optimal Parameters:}  
Finally, the optimal set of potential parameters $\mu^\ast$, $\tilde{V}_A^\ast$, and $\tilde{V}_R^\ast$ are obtained.  
The values $\tilde{V}_A^\ast$ and $\tilde{V}_R^\ast$ correspond to the outputs of the trained networks $F_1$ and $F_2$ at $\mu = \mu^\ast$.
\end{enumerate}
This inversion procedure is carried out independently for $^1S_0$ state of neutron-neutron (n-n), neutron-proton (n-p), and proton-proton (p-p) scattering systems, each using their respective expected phase shift datasets. This ensures that the estimated potential parameters are consistent with the underlying scattering dynamics of each system.
\section{Results and Discussion}
\subsection{Input Data}
For training the neural networks $F_1$ and $F_2$, we generated synthetic input data corresponding to the Malfliet--Tjon (MT) potential parameters, $V_A$, $V_R$, and $\mu$, as described in the methodology. The potential parameters were uniformly sampled within the ranges: $V_A \in [-100,\,100] \ \mathrm{fm^{-2}}$, $V_R \in [-300,\,300] \ \mathrm{fm^{-2}}$, 
$\mu \in [0.01,\,4] \ \mathrm{fm^{-1}}$.
For each sampled configuration, the phase function equation was solved for $\ell = 0$ using the fifth-order Runge-Kutta (RK-5)  method \cite{Sharma2014} to compute the phase shifts at the laboratory energies:\[
E_{\mathrm{lab}} = [1,\,5,\,10,\,25,\,50,\,100,\,150,\,200,\,250,\,300,\,350] \ \mathrm{MeV}.\] A total of 10,000 unique potential profiles were generated for the n-n, n-p, and p-p scattering systems. These computed phase shifts were used as the input features for training the models, while the target outputs correspond to the potential parameters.
\subsection{Training and Validation Performance}

The generated dataset was divided into three subsets: $70\%$ for training, $20\%$ for validation, and $10\%$ for testing. Both models, $F_{1}$ and $F_{2}$, share an identical architecture consisting of three fully connected hidden layers with $128$, $64$, and $64$ neurons, respectively. The Rectified Linear Unit (ReLU) activation function was used after each hidden layer to introduce non-linearity.
The models were trained using the Adam optimizer with a learning rate of $10^{-3}$, and the Mean Squared Error (MSE) was employed as the loss function. Training was performed for a total of 2000 epochs,during which the loss decreased steadily. The training and validation curves show smooth convergence, indicating that the models effectively learned the mapping between the phase shifts and the potential parameters without significant overfitting. The testing dataset was used for final performance evaluation, confirming the generalization capability of the proposed two-stage network framework.

\begin{figure}[h!]
    \centering
    \begin{subfigure}[t]{0.48\textwidth}
        \centering
        \includegraphics[width=\textwidth]{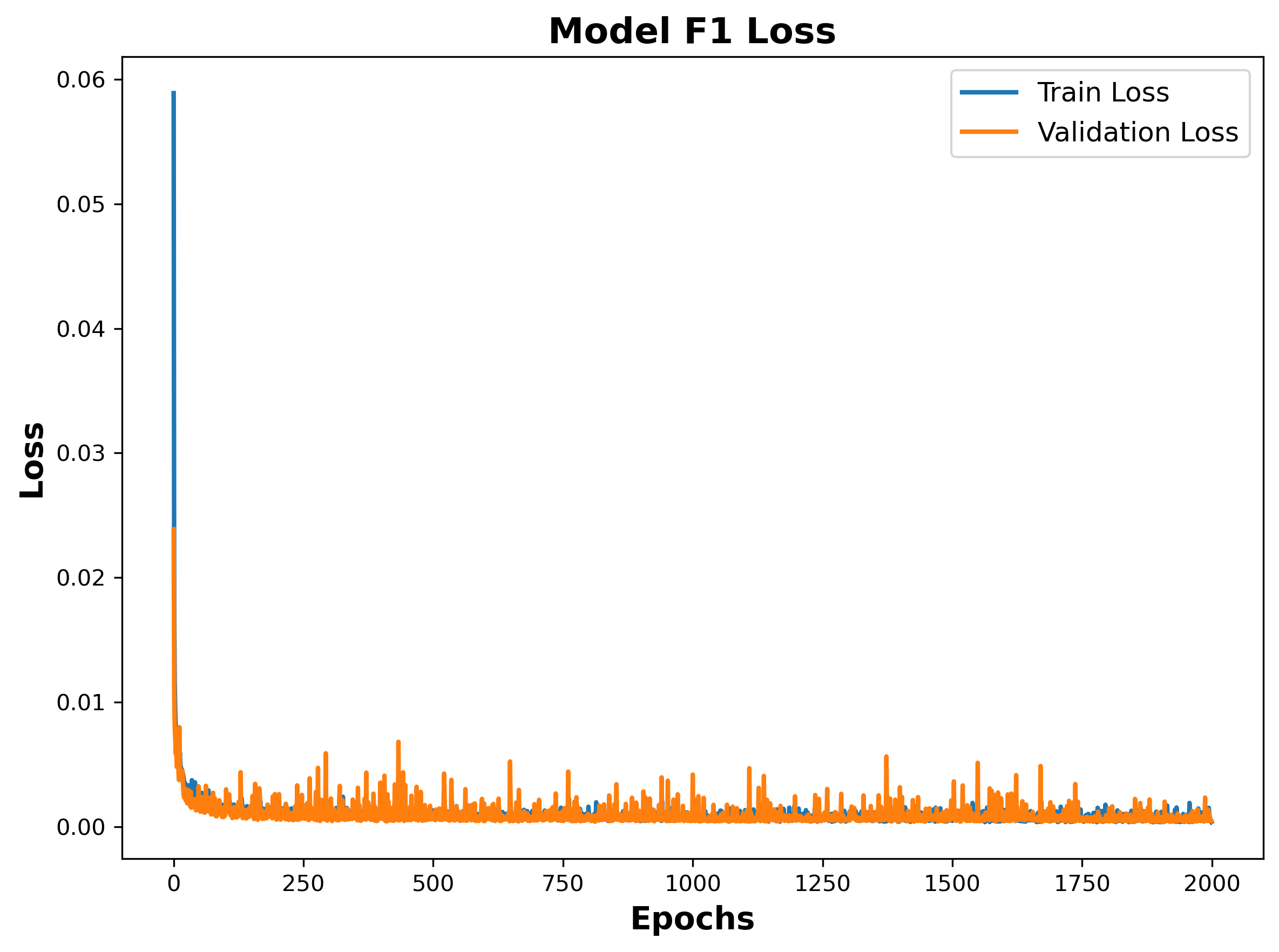}
        \caption{Training and validation loss for $F_{1}$ on neutron-neutron scattering.}
        \label{fig:F1_nn_loss}
    \end{subfigure}
    \hfill
    \begin{subfigure}[t]{0.48\textwidth}
        \centering
        \includegraphics[width=\textwidth]{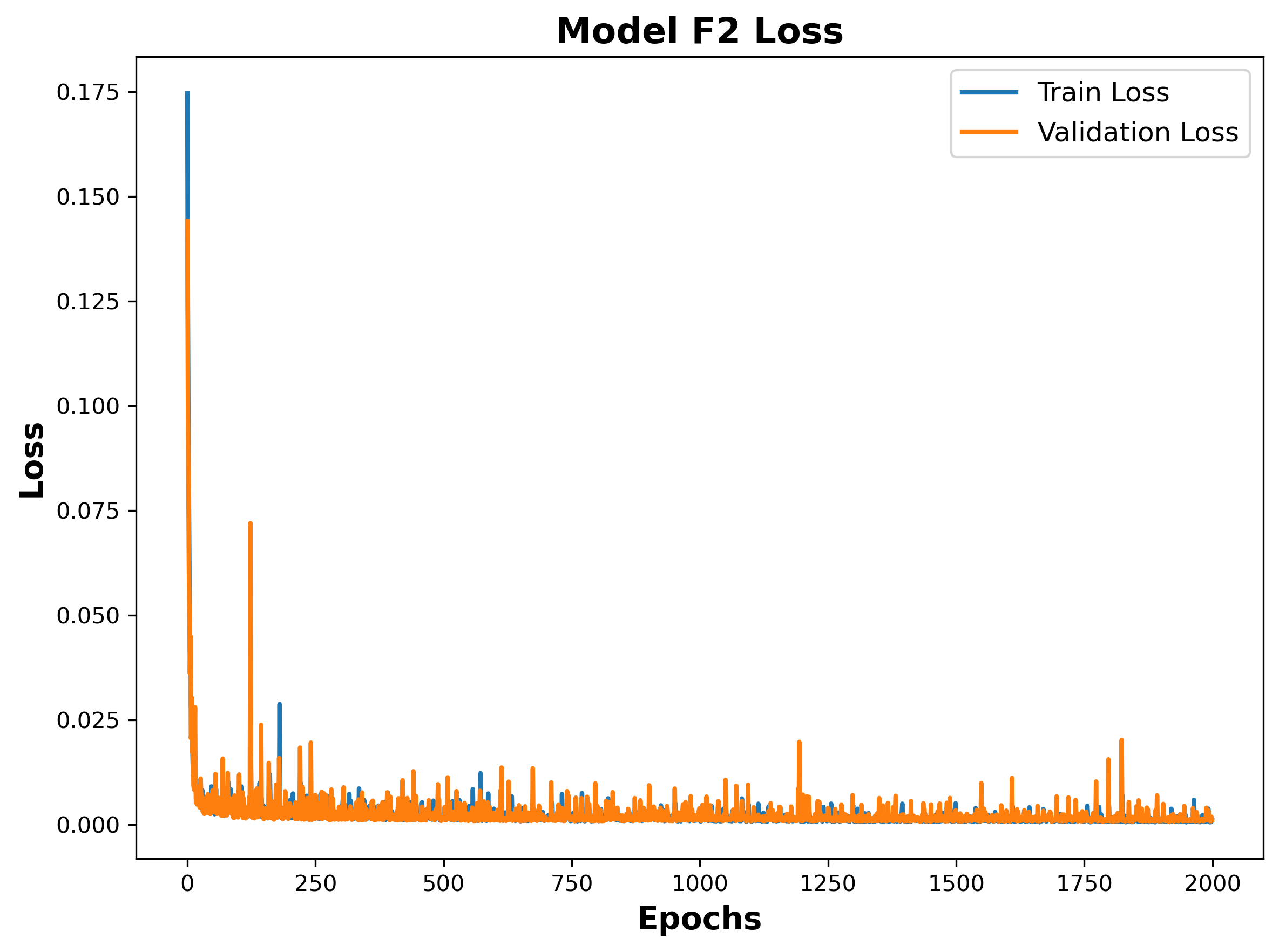}
        \caption{Training and validation loss for $F_{2}$ on neutron-neutron scattering.}
        \label{fig:F2_nn_loss}
    \end{subfigure}
    \caption{Training and validation loss curves for n-n scattering.}
    \label{fig:nn_loss}
\end{figure}

\begin{figure}[h!]
    \centering
    \begin{subfigure}[t]{0.48\textwidth}
        \centering
        \includegraphics[width=\textwidth]{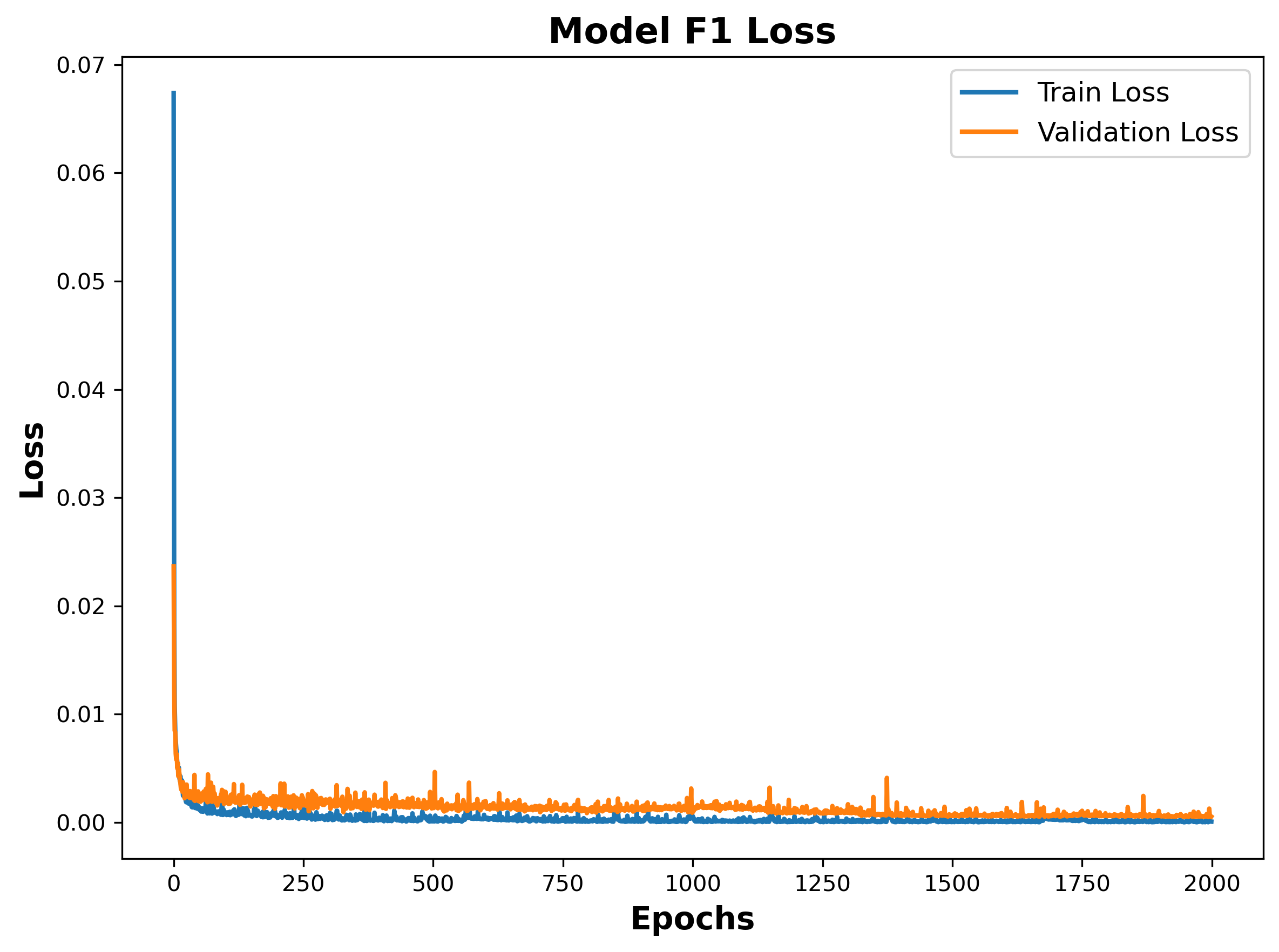}
        \caption{Training and validation loss for $F_{1}$ on neutron-proton scattering.}
        \label{fig:F1_np_loss}
    \end{subfigure}
    \hfill
    \begin{subfigure}[t]{0.48\textwidth}
        \centering
        \includegraphics[width=\textwidth]{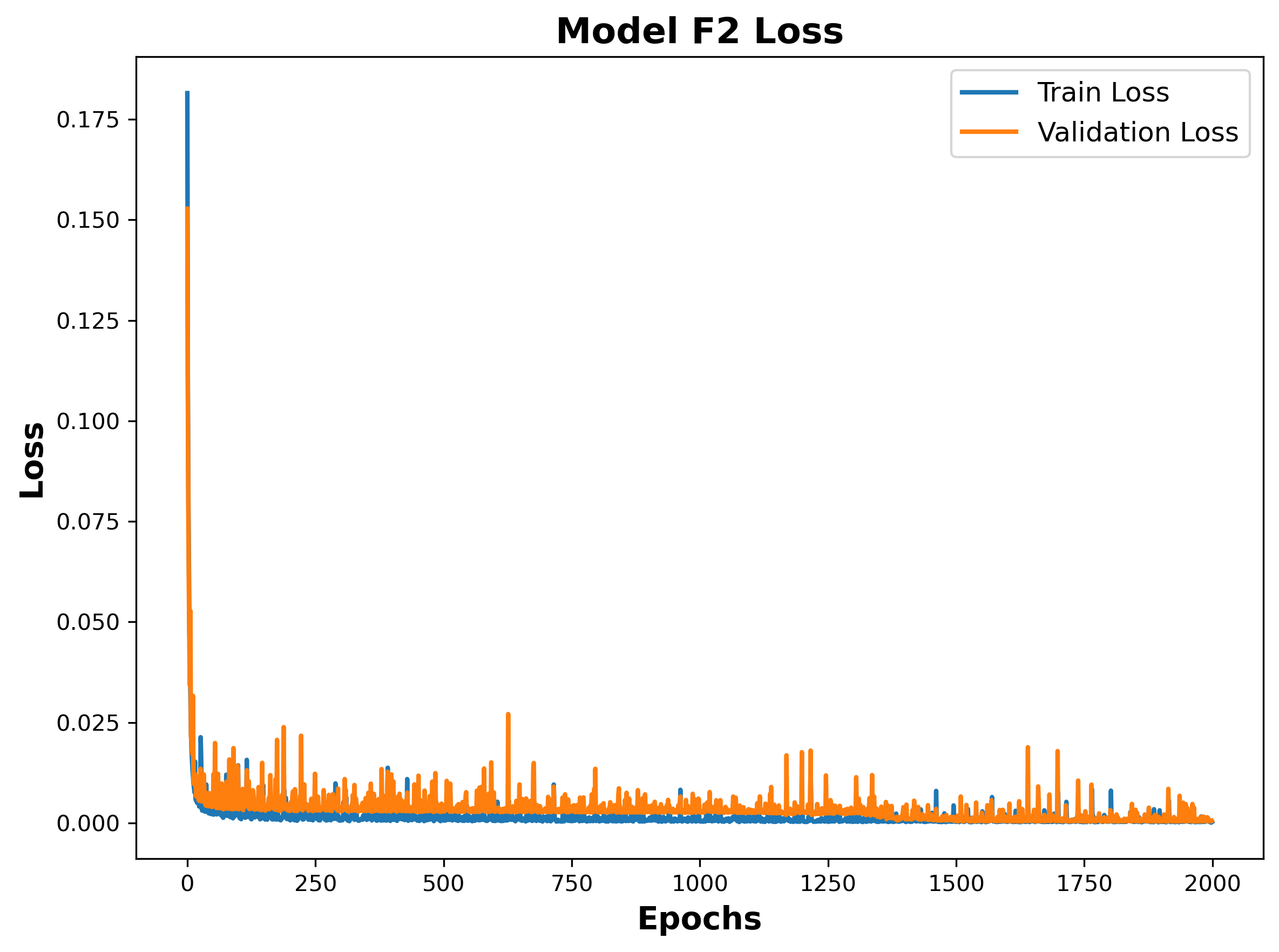}
        \caption{Training and validation loss for $F_{2}$ on neutron-proton scattering.}
        \label{fig:F2_np_loss}
    \end{subfigure}
    \caption{Training and validation loss curves for n-p scattering.}
    \label{fig:np_loss}
\end{figure}

\begin{figure}[h!]
    \centering
    \begin{subfigure}[t]{0.48\textwidth}
        \centering
        \includegraphics[width=\textwidth]{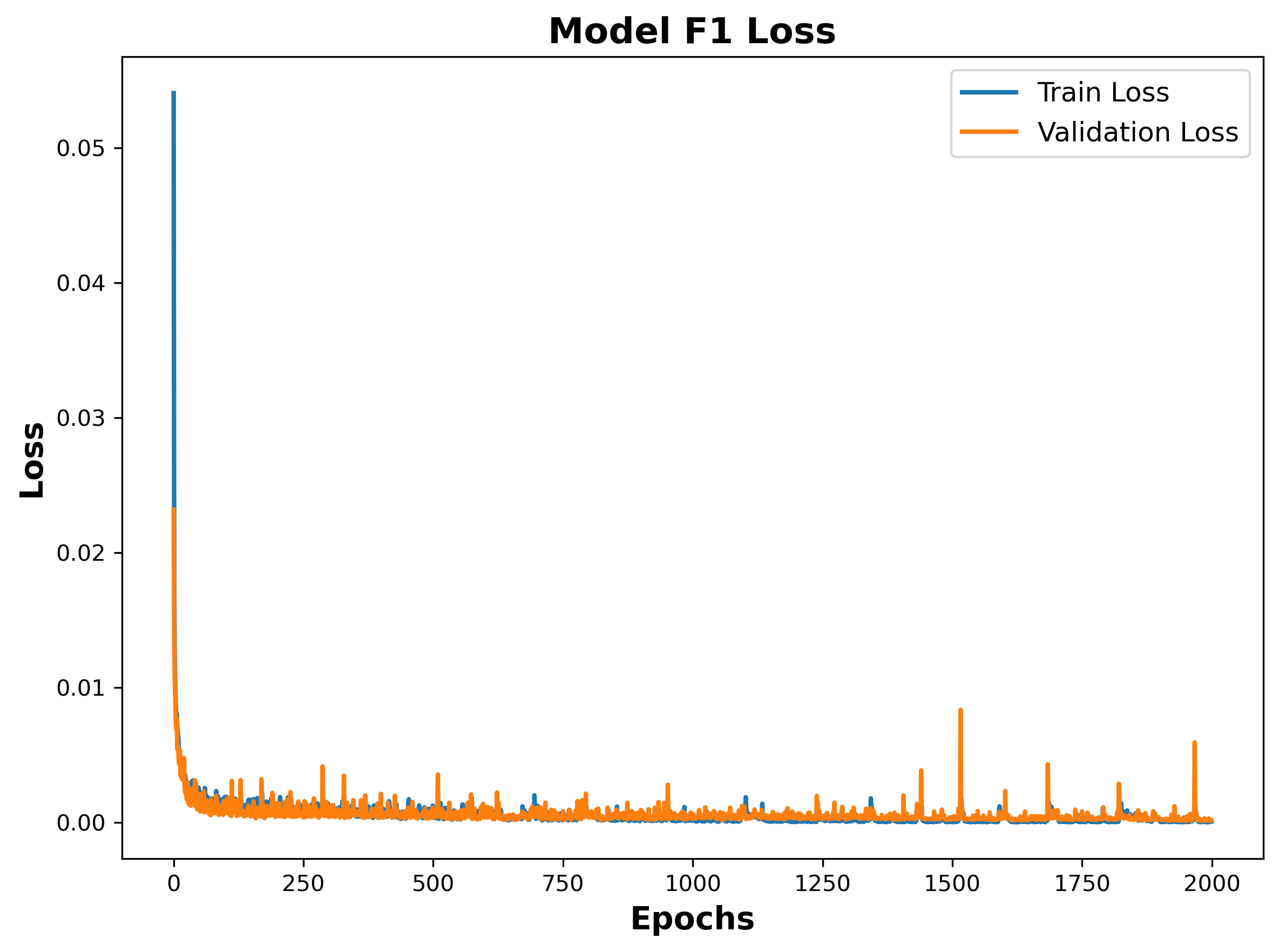}
        \caption{Training and validation loss for $F_{1}$ on proton-proton scattering.}
        \label{fig:F1_pp_loss}
    \end{subfigure}
    \hfill
    \begin{subfigure}[t]{0.48\textwidth}
        \centering
        \includegraphics[width=\textwidth]{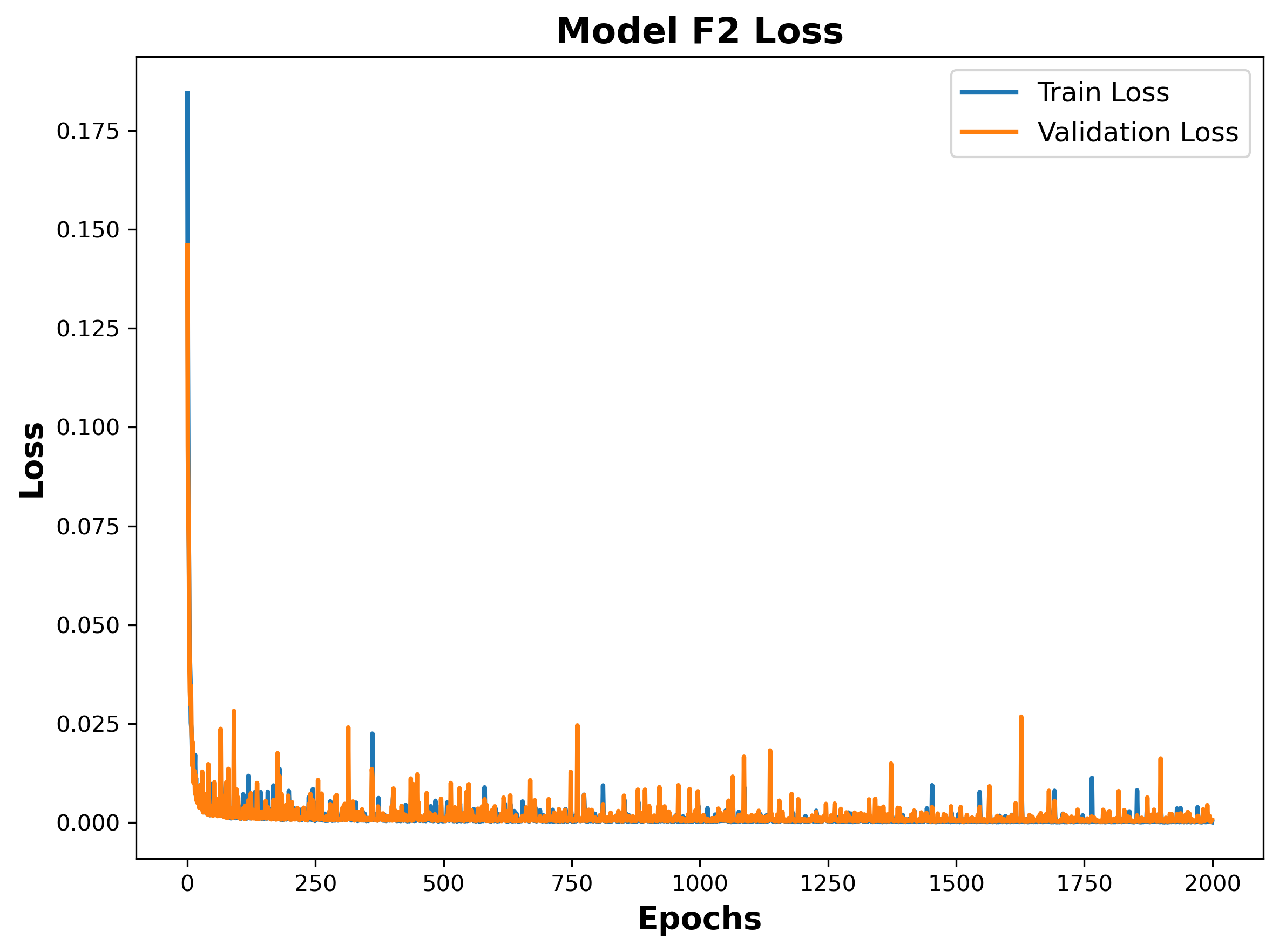}
        \caption{Training and validation loss for $F_{2}$ on proton-proton scattering.}
        \label{fig:F2_pp_loss}
    \end{subfigure}
    \caption{Training and validation loss curves for p-p scattering.}
    \label{fig:pp_loss}
\end{figure}

Figures~\ref{fig:nn_loss}--\ref{fig:pp_loss} show the training and validation loss curves for $F_{1}$ and $F_{2}$ across all three scattering systems: neutron-neutron (n-n), neutron-proton (n-p), and proton-proton (p-p). Each figure presents the loss curves for both models side by side for better comparison. In all three scattering cases, the training and validation losses decrease smoothly, and the validation curves follow the training curves closely, indicating stable learning without overfitting.

To quantitatively evaluate the performance, Table~\ref{tab:val_loss} summarizes the final validation losses for $F_{1}$ and $F_{2}$ across all three scattering systems.
\begin{table}[h!]
\centering
\caption{Final validation losses for $F_1$ and $F_2$ models across all scattering systems.}
\begin{tabular}{|c|c|c|}
\hline
\textbf{Scattering System} & \textbf{Validation Loss ($F_1$)} & \textbf{Validation Loss ($F_2$)} \\
\hline
n-n & $7.00 \times 10^{-4}$ & $6.59 \times 10^{-3}$ \\
n-p & $5.49 \times 10^{-4}$ & $5.22 \times 10^{-4}$ \\
p-p & $6.20 \times 10^{-4}$ & $6.29 \times 10^{-4}$ \\
\hline
\end{tabular}
\label{tab:val_loss}
\end{table}

\noindent
From Table~\ref{tab:val_loss}, it is evident that:
\begin{itemize}
    \item For n-p and p-p scattering, both models achieve very low validation losses $(<7\times10^{-4})$, indicating excellent generalization.
    \item In n-n scattering, Model $\textbf{F}_{1}$ outperforms Model $\textbf{F}_{2}$, which is expected since estimating $\tilde{V_A}$ is less complex than predicting $\tilde{V_R}$.
    \item The smooth convergence of both training and validation losses across all scattering systems demonstrates the stability and robustness of the proposed framework.
\end{itemize}

Thus, the proposed two-stage neural network framework effectively learns the mapping between the phase shifts and the potential parameters $V_A$ and $V_R$. The combination of visual convergence patterns and quantitative validation results confirms the stability, accuracy, and generalization capability of the models.

\subsubsection*{Prediction Performance}
To evaluate the prediction capability of the proposed neural network framework, we tested the models $F_{1}$ and $F_{2}$ on an independent test dataset. For the sake of simplicity, we plotted $N = 30$ random data points from the testing set and plotted the estimated potential parameters $V_A$ and $V_R$ against the data index $N$ for all three scattering cases. In these plots, the blue curves represent the true values of the model parameters obtained from the sampled dataset, while the red curves represent the predicted (estimated) values generated by the trained models. By comparing the two curves, we can analyze how accurately the models are able to estimate the potential parameters based on the given phase shift information.

\begin{figure}[h!]
    \centering
    \includegraphics[width=0.7\textwidth]{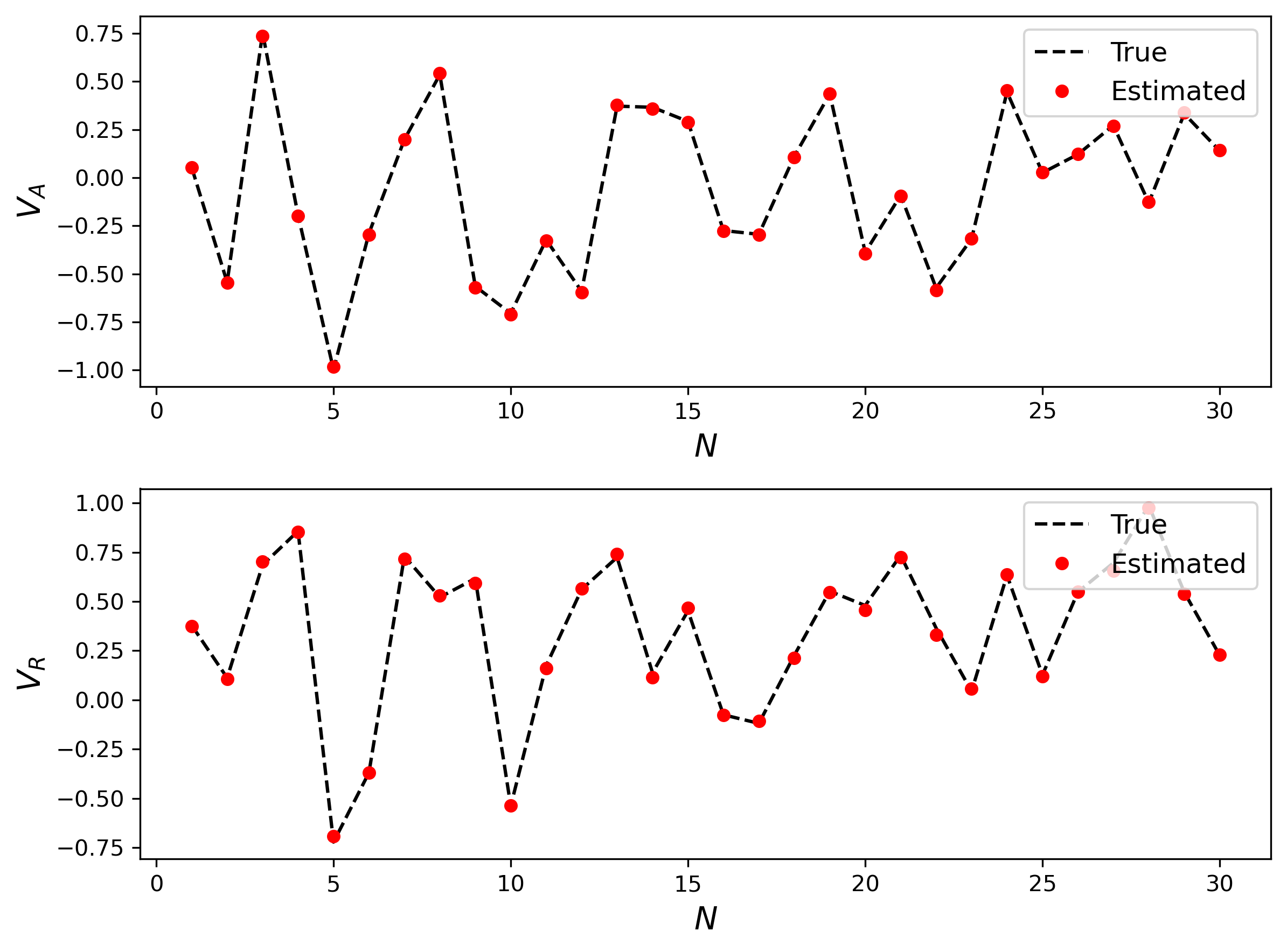}
    \caption{Prediction performance of $V_A$ and $V_R$ for neutron-neutron (n-n) scattering on $N=30$ testing data points. The blue curve shows the true values and the red curve represents the predicted values.}
    \label{fig:nn_pred}
\end{figure}

\begin{figure}[h!]
    \centering
    \includegraphics[width=0.7\textwidth]{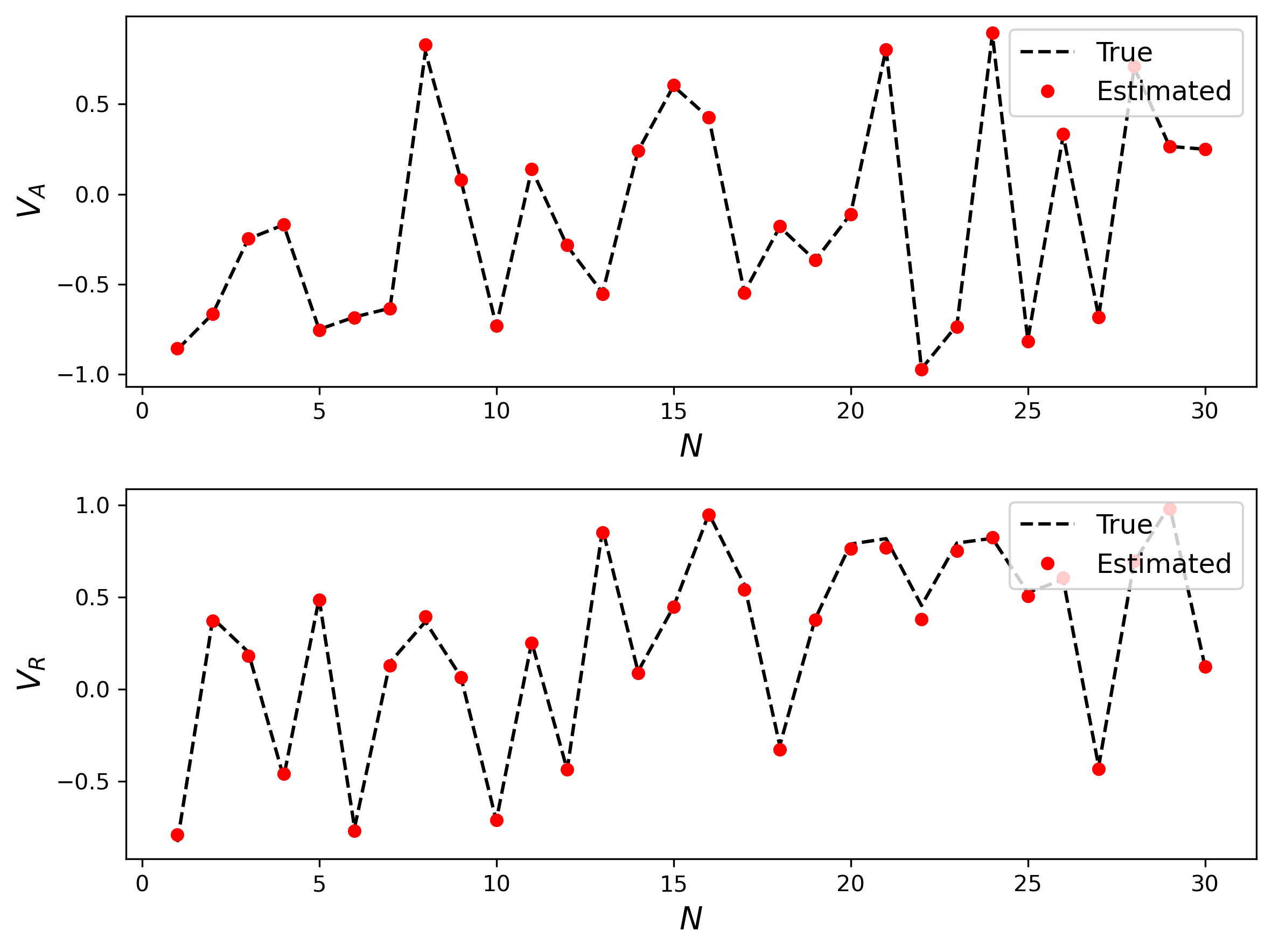}
    \caption{Prediction performance of $V_A$ and $V_R$ for neutron-proton (n-p) scattering on $N=30$ testing data points. The blue curve shows the true values and the red curve represents the predicted values.}
    \label{fig:np_pred}
\end{figure}

\begin{figure}[h!]
    \centering
    \includegraphics[width=0.7\textwidth]{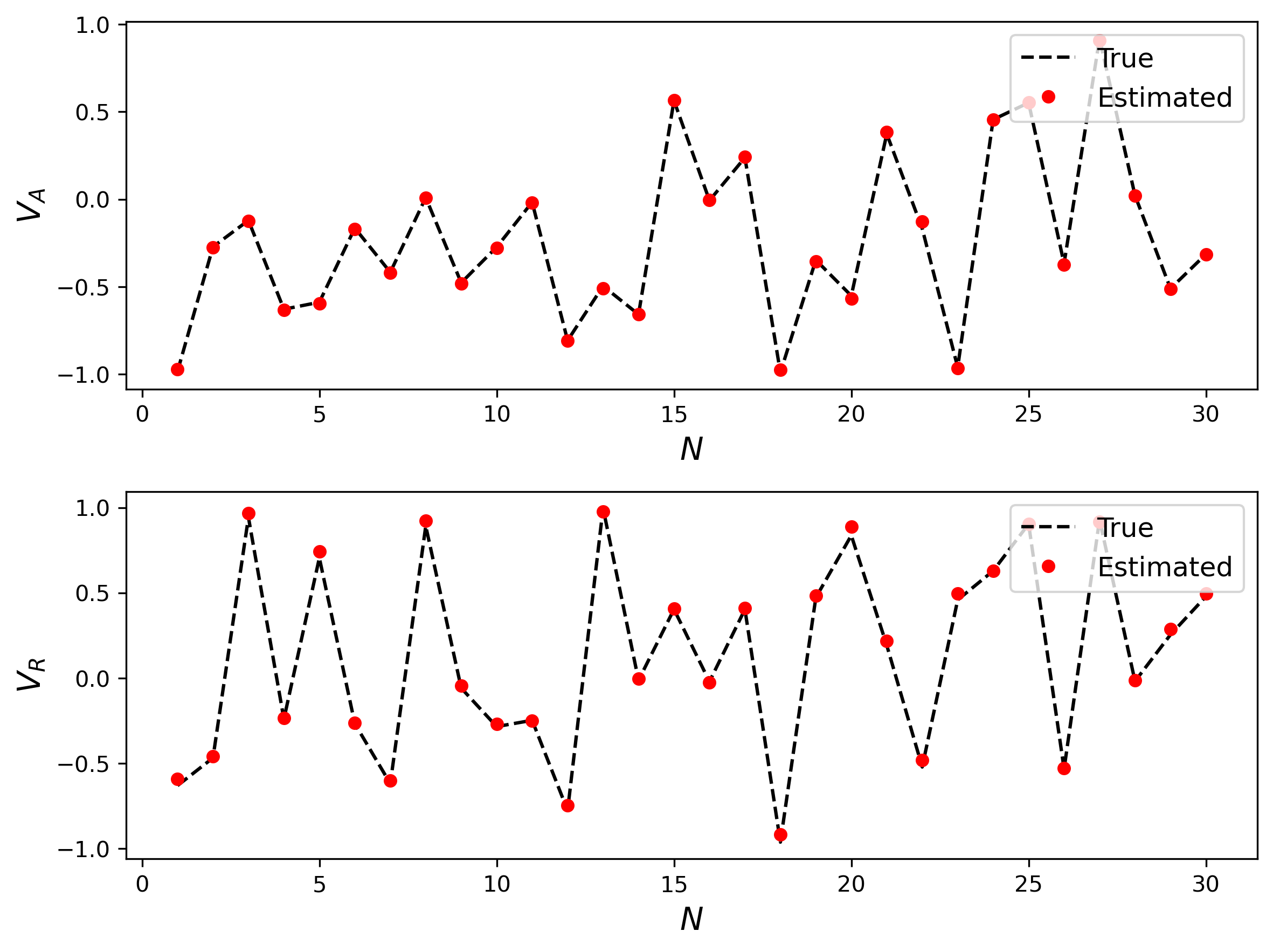}
    \caption{Prediction performance of $V_A$ and $V_R$ for proton-proton (p-p) scattering on $N=30$ testing data points. The blue curve shows the true values and the red curve represents the predicted values.}
    \label{fig:pp_pred}
\end{figure}

Figures~\ref{fig:nn_pred}, \ref{fig:np_pred}, and \ref{fig:pp_pred} illustrate the prediction performance for the three scattering scenarios. It can be observed that the predicted values of both $V_A$ and $V_R$ are in very close agreement with the actual values, showing minimal deviation across all $N = 30$ test points. This indicates that the proposed model effectively captures the underlying non-linear relationship between the phase shifts and the potential parameters.

From these figures, it is evident that the trained network,  provides highly accurate predictions of the potential parameters. The closeness of the predicted and true values demonstrates the reliability, stability, and generalization capability of the model. The consistent performance across the n-n, n-p, and p-p scattering systems confirms that the proposed approach successfully learns the mapping from phase shift data to the underlying potential parameters.
\subsection{Estimation of Optimal Potential Parameters}
\begin{figure}[h!]
    \centering
    \includegraphics[width=0.8\linewidth]{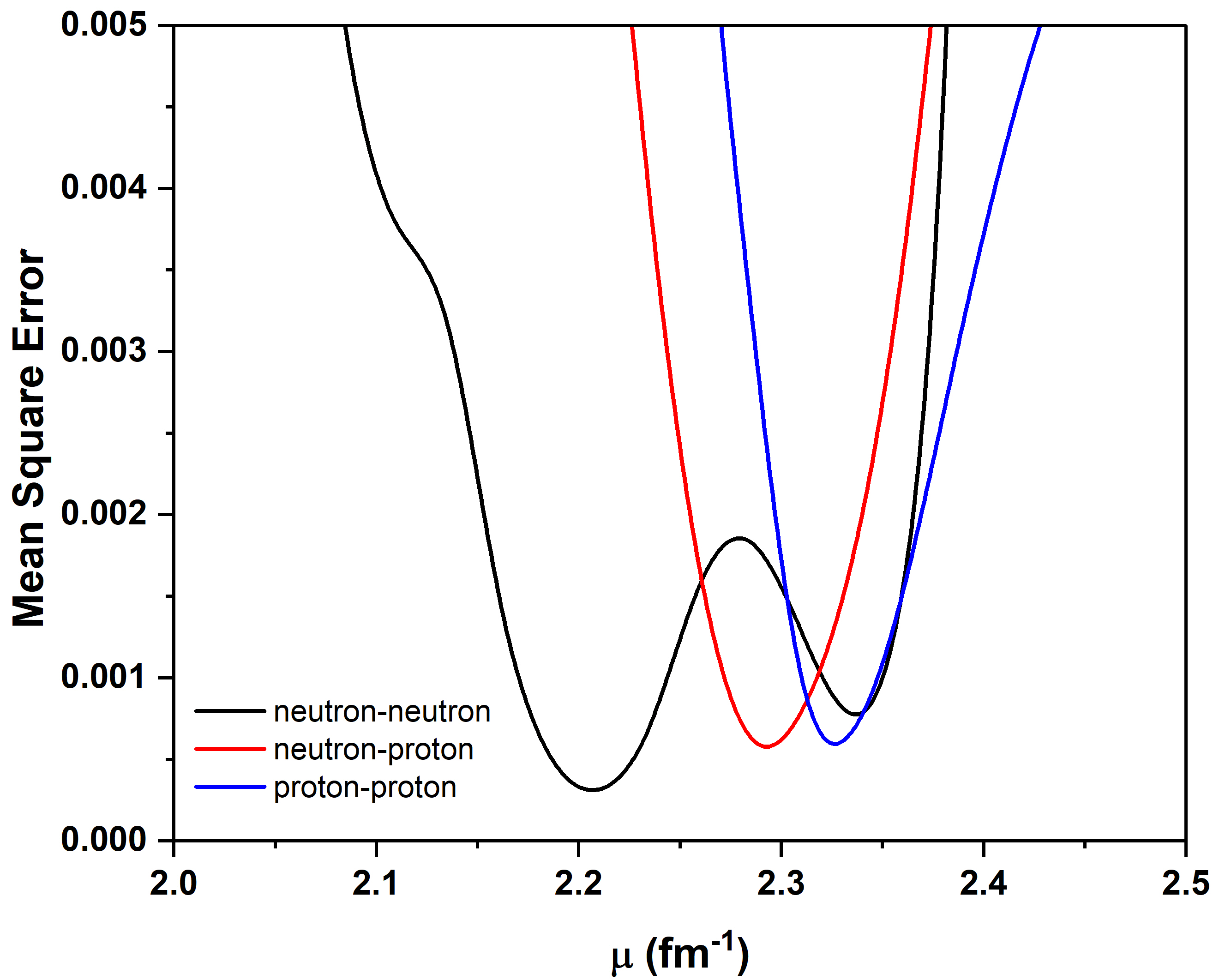}
    \caption{Variation of MSE with $\mu$ for neutron-neutron (black), neutron-proton (red), 
    and proton-proton (blue) scattering systems in the range $2.0$-$2.5$ fm$^{-1}$. 
    The minima correspond to the optimal $\mu$ values, which are then used to obtain 
    the corresponding $V_A$ and $V_R$ from the trained neural networks.}
    \label{fig:mse_comparison}
\end{figure}

To estimate the optimal potential parameters $\left(V_A^*, V_R^*, \mu^*\right)$ for the n-n, n-p, and p-p scattering systems, we employed the inversion framework described 
in Section~2.3. In this procedure, the range parameter $\mu$ was varied within the 
interval $[0.01, 4]~\mathrm{fm^{-1}}$. For each trial value of $\mu$, the trained neural networks $F_1$ and $F_2$ were used to predict the corresponding potential strengths $V_A^*$ and $V_R^*$. To identify the optimal $\mu^*$, we plotted the variation of MSE with $\mu$ for each scattering system in the zoomed range $2.0 \leq \mu \leq 2.5~\mathrm{fm^{-1}}$, as shown in Figure~\ref{fig:mse_comparison}. From this figure, it is evident that the MSE curve decreases sharply, reaches a distinct minimum, and then increases again. These minima correspond to the optimal values of $\mu^*$ for the three scattering systems:\[
\mu_{\mathrm{opt}}^{(n-n)} = 2.210~\mathrm{fm^{-1}}, \quad
\mu_{\mathrm{opt}}^{(n-p)} = 2.285~\mathrm{fm^{-1}}, \quad
\mu_{\mathrm{opt}}^{(p-p)} = 2.335~\mathrm{fm^{-1}}.
\]
Once the optimal $\mu$ is identified, the corresponding values of $V_A^*$ and $V_R^*$ are obtained directly from the trained neural networks $F_1$ and $F_2$ at this specific $\mu_{\mathrm{opt}}$. Thus, for each scattering system, the complete set of optimal parameters $\left(V_A^*, V_R^*, \mu^*\right)$ is determined.
The estimated optimal parameters for the three scattering systems are 
given in Table~\ref{tab:opt_params}.
\begin{table}[h!]
\centering
\caption{Optimal potential parameters obtained for $^1S_0$ state of n-n, n-p, and p-p scattering systems.}
\begin{tabular}{|c|c|c|c|}
\hline
\textbf{Scattering System} & \boldmath$V_A^*$ \textbf{(fm$^{-2}$)} & 
\boldmath$V_R^*$ \textbf{(fm$^{-2}$)} & \boldmath$\mu^*$ \textbf{(fm$^{-1}$)} \\
\hline
n-n & 29.731 & 123.411 & 2.210 \\
n-p & 30.344 & 123.095 & 2.285 \\
p-p & 33.797 & 147.367 & 2.335 \\
\hline
\end{tabular}
\label{tab:opt_params}
\end{table}
From Table~\ref{tab:opt_params}, it is observed that the $p$-$p$ scattering system 
requires a significantly higher repulsive strength $V_R$ compared to the $n$-$n$ and $n$-$p$ systems due to the additional Coulomb repulsion between protons. In contrast, the $n$-$n$ and $n$-$p$ systems show relatively similar parameter values, which is consistent with the expected symmetry of nuclear interactions.
\subsection{Construction of Inverse Potentials and Phase Shift Analysis}
\begin{figure}[h!]
\centering
\includegraphics[width=0.75\linewidth]{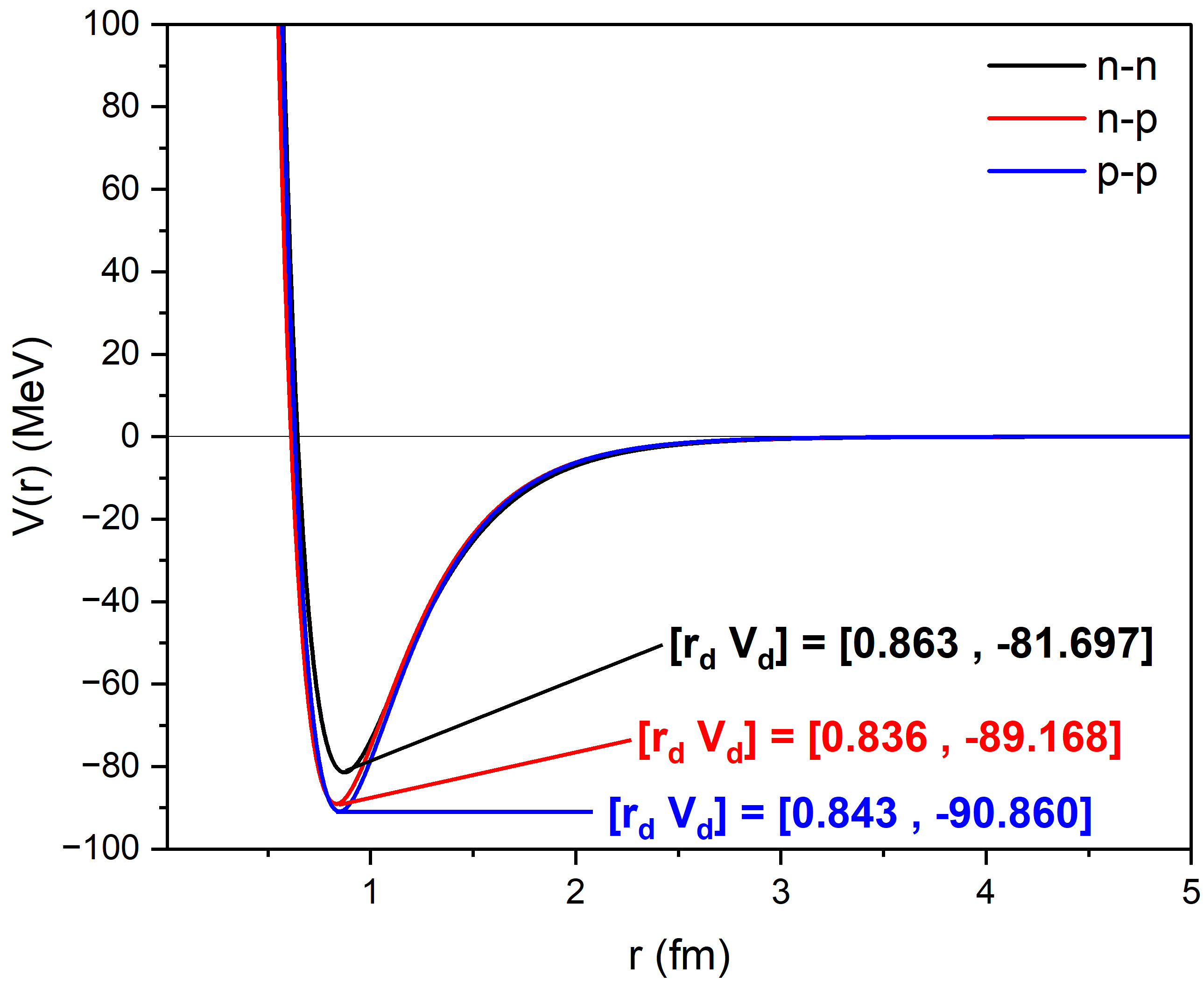}
\caption{Constructed inverse potentials $V(r)$ in MeV for the n-n, n-p, and p-p scattering systems using the optimal parameters from Table~\ref{tab:opt_params}.}
\label{pot}
\end{figure}

Using the optimal parameters listed in Table~\ref{tab:opt_params}, the inverse potentials for the n-n, n-p, and p-p scattering systems were constructed. These potentials were obtained by substituting the optimized parameter set \(\left(V_A^*, V_R^*, \mu^*\right)\) into the Malfliet--Tjon potential defined in Eq.~(6). The potential values obtained from this substitution were initially expressed in units of \(\mathrm{fm}^{-2}\). To convert these values into energy units (\(\mathrm{MeV}\)), we multiplied them by the factor \(\dfrac{\hbar^2}{2m}\), where m denotes the reduced mass of the two-nucleon system. The final constructed potentials \(V(r)\), expressed in \(\mathrm{MeV}\) as a function of the radial distance \(r\), are presented in Fig.~\ref{pot}.
It is observed that all three constructed potentials exhibit the characteristic features of nucleon-nucleon interactions, namely a short-range repulsive core followed by an intermediate-range attractive well. Among the three, the p--p potential shows a slightly stronger repulsion at short distances due to the additional Coulomb interaction, whereas the n-n and n-p potentials remain relatively close in magnitude, which is consistent with the approximate charge symmetry of nuclear forces \cite{Naghdi2014}.
From Fig.~\ref{pot}, it is observed that the depth of the potential, \(V_d\), for the n-n system is found to be \(-81.697~\mathrm{MeV}\) at a corresponding  equilibrium distance of \(r_d = 0.863~\mathrm{fm}\). Similarly, for the n-p system, the potential depth is \(-89.168~\mathrm{MeV}\) at \(r_d = 0.836~\mathrm{fm}\), while for the p-p system, the potential depth reaches \(-90.860~\mathrm{MeV}\) at \(r_d = 0.843~\mathrm{fm}\).
These results indicate that the n-p system exhibits the deepest attractive well among the three, which is consistent with the stronger binding tendency in neutron-proton interactions compared to neutron-neutron scattering. The slightly shallower well for the n-n system reflects the weaker attractive interaction due to the absence of isospin mixing effects. Furthermore, the p-p potential shows a depth comparable to the n-p case but at a slightly larger equilibrium distance, which can be attributed to the additional repulsive contribution from the Coulomb interaction between protons. Overall, the observed variations in \(V_d\) and \(r_d\) across the three systems are consistent with the expected isospin dependence and charge-symmetry breaking in nucleon-nucleon interactions \cite{Naghdi2014,PhysRevC.63.024001}.
\subsubsection*{Effect of Local vs. Global Minimum in neutron-neutron (n-n) Scattering}
\begin{figure}[h!]
    \centering
    \includegraphics[width=1.0\textwidth,height=0.7\textheight,keepaspectratio]{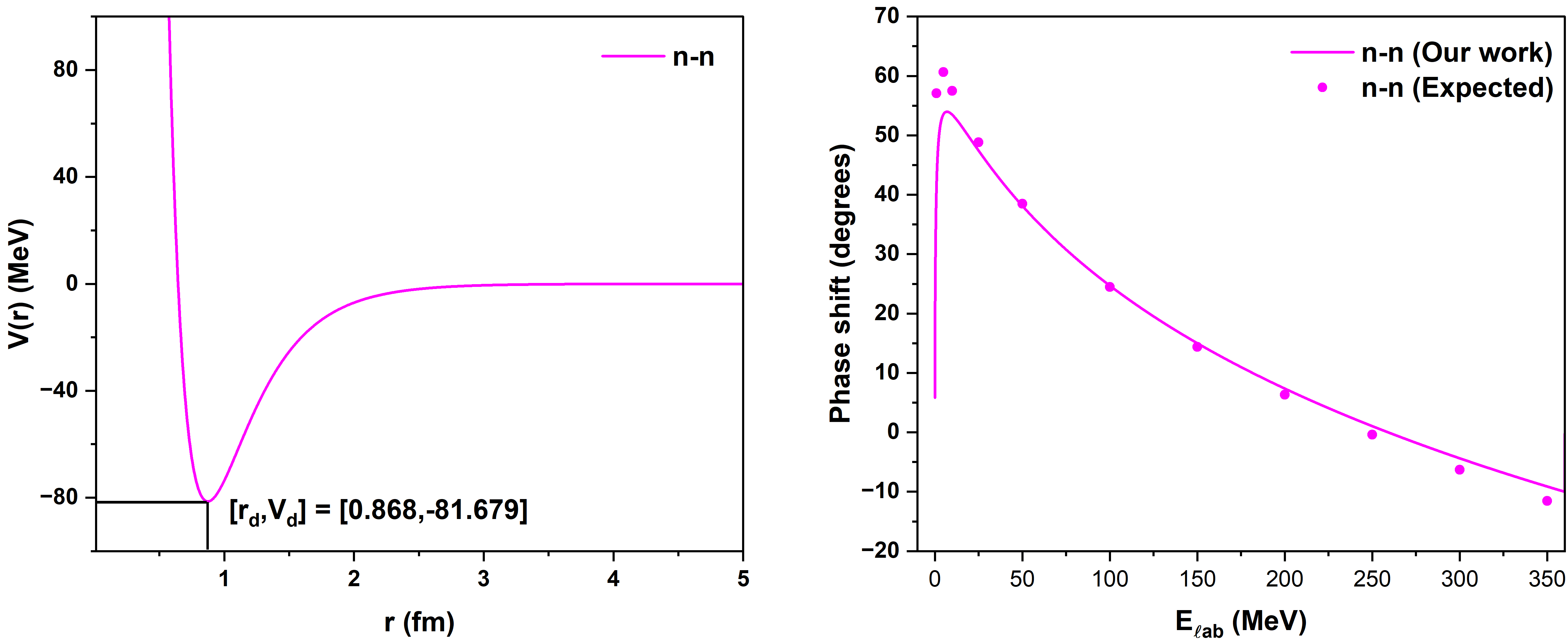}
    \caption{Constructed inverse potential and corresponding scattering phase shifts for neutron-neutron scattering obtained by taking $\mu = 2.335~\mathrm{fm^{-1}}$.}
    \label{l_g}
\end{figure}

From Fig.~\ref{fig:mse_comparison}, it is observed that for n-n 
scattering, two distinct minima occur at $\mu = 2.210~\mathrm{fm^{-1}}$ (global) and $\mu = 2.335~\mathrm{fm^{-1}}$ (local). To investigate the effect of selecting a local minimum instead of the global minimum, we consider the case where $\mu = 2.335~\mathrm{fm^{-1}}$ is taken as the optimal parameter. Using this value of $\mu$, the corresponding attractive and repulsive parameters are obtained from models $F_1$ and $F_2$, respectively. Based on these parameters, the inverse potential is constructed, and the corresponding phase shifts are computed, as shown in Fig.~\ref{l_g}.
From this figure, it is observed that although the depth of the potential and the equilibrium distances vary only slightly when using $\mu = 2.335~\mathrm{fm^{-1}}$, the phase shifts exhibit significant deviations, particularly at low energies. 
At these energies, the predicted phase shifts fail to reproduce the expected behavior accurately.  This analysis demonstrates that selecting a local minimum 
can reduce the predictive accuracy of the interaction potential. 
Therefore, to obtain a high-precision description of the nucleon-nucleon interaction, it is crucial to select the global minimum. 
Consequently, $\mu = 2.210~\mathrm{fm^{-1}}$ is identified as the 
optimal parameter, providing the most accurate interaction potential 
and yielding phase shifts in excellent agreement with the expected results.

\begin{figure}[h!]
\centering
\includegraphics[width=1.0\textwidth,height=0.7\textheight,keepaspectratio]{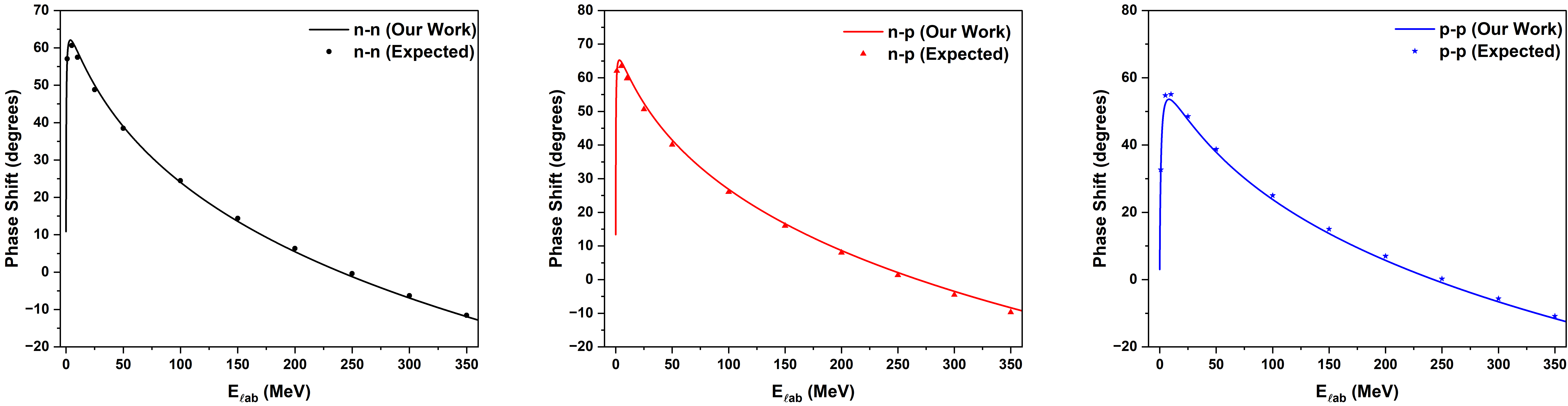}
\caption{Comparison of predicted and expected phase shifts from Wiringa \textit{et al.}~\cite{PhysRevC.51.38} for n-n, n-p, and p-p scattering systems as a function of laboratory energy. }
\label{sps}
\end{figure}

So, using the obtained inverse potentials from the \textit{``global minima"}, we calculated the scattering phase shifts for the n-n, n-p, and p-p systems. The computed phase shifts are plotted against the laboratory energy and compared with the expected phase shifts reported by Wiringa \textit{et al.}~\cite{PhysRevC.51.38}, as shown in Fig.~\ref{sps}. From Fig.~\ref{sps}, it is observed that the predicted phase shifts match closely with the expected values over the considered energy range, demonstrating the effectiveness of the proposed inversion approach and validating the accuracy of the constructed potentials.

\section{Conclusions}
In this work, we have presented a comprehensive physics-guided neural network (PGNN) framework for constructing nucleon-nucleon inverse potentials. By integrating the Phase Function Method with a two-stage neural network inversion procedure, the approach enables the systematic estimation of Malfliet-Tjon (MT) potential parameters for neutron-neutron (n-n), neutron-proton (n-p), and proton-proton (p-p) scattering systems. Our results demonstrate that the PGNN framework effectively optimizes potential parameters and constructs inverse potentials that are consistent with established features of nucleon-nucleon interactions. Specifically, the n-p potential exhibits the most pronounced attractive well, while the p-p potential shows stronger short-range repulsion, consistent with the Coulomb interaction between protons. Additionally, the simulated phase shifts closely match with the expected data, indicating that the inversion process is both accurate and physically reliable.\\
This study underscores the significant advantages of combining machine learning techniques with physics-based modeling. The PGNN approach enhances the stability of the inversion process, improves predictive accuracy, and reduces computational cost relative to traditional inversion techniques. Beyond the immediate results, this framework provides a flexible foundation for addressing more complex nuclear physics problems. Future research will focus on extending the methodology to higher partial waves and three-body nuclear systems, thereby broadening its applicability to a wider range of nuclear structure and reaction studies.
\\ Overall, this framework illustrates a promising pathway for leveraging neural networks in tandem with theoretical physics to tackle challenging inverse scattering problems, bridging the gap between data-driven methods and fundamental physical principles.
\\
\textbf{Acknowledgments}
A. Awasthi acknowledges financial support provided by Department of Science and Technology (DST), Government of India vide Grant No. DST/INSPIRE Fellowship/2020/IF200538. 
\\ 
\textbf{Author Declaration} 
The authors declare that they have no conflict of interest.

\end{document}